%% file: SplitDraftCR.tex
\pdfoutput=1
 
\documentclass[conference]{IEEEtran}
\usepackage[font=small]{caption}
\usepackage{color}
\usepackage{amsmath}
\usepackage{mathtools}
\usepackage{amsmath}
\usepackage{amssymb}
\usepackage{cite}
\usepackage{array}
\usepackage{amsthm}
\usepackage{amsmath}
\usepackage{tabulary}
\usepackage{multirow}
\usepackage{hyperref}
\usepackage{subfig}
\usepackage{enumerate}
\IEEEoverridecommandlockouts
\usepackage{pifont}
\usepackage{epsfig}
\usepackage{graphicx}
\usepackage{url}
\usepackage{amssymb}
\usepackage{paralist} %
\usepackage{algorithm}
\usepackage{algpseudocode}
\usepackage{pifont}

\graphicspath{ {./figs/} }

\include{notation}

\begin{document}
\title{Optimal Traffic Aggregation in  Multi-RAT Heterogeneous Wireless Networks}

\author{
\IEEEauthorblockN{Sarabjot Singh, Shu-ping Yeh, Nageen Himayat, and Shilpa Talwar}
\IEEEauthorblockA{Intel Corportation, Santa Clara, CA, USA
    \\ \{sarabjot.singh\}, \{shu-ping.yeh\}, \{nageen.himayat\}, \{shilpa.talwar\}@intel.com}
}
\maketitle

\begin{abstract}
Traffic load balancing and radio resource management is key to harness the dense and increasingly heterogeneous deployment of next generation ``$5$G" wireless infrastructure. Strategies for aggregating  user traffic  from across multiple radio access technologies (RATs) and/or access points (APs) would be crucial in such heterogeneous networks (HetNets), but are not well investigated. In this paper, we develop a low complexity solution for maximizing an $\alpha$-optimal network utility leveraging the multi-link aggregation (simultaneous connectivity to multiple RATs/APs) capability of users in the network.  The network utility maximization formulation has maximization of sum rate ($\alpha=0$), maximization of minimum rate ($\alpha \to \infty$), and proportional fair ($\alpha=1$)  as its special cases.  A closed form  is also developed for the special case where a user aggregates traffic from at most two APs/RATs, and hence can be applied to practical scenarios like LTE-WLAN aggregation (LWA) and LTE dual-connectivity solutions.  It is shown that the required objective  may also be realized through a decentralized implementation requiring a series of message exchanges between the users and network. Using comprehensive system level simulations, it is shown that optimal leveraging of multi-link aggregation  leads to substantial throughput  gains  over   single RAT/AP selection techniques.
\end{abstract}

\section{Introduction}
Drastically increasing demand for wireless data and devices \cite{cisco} has led to an increasing requirement for both peak  rates and area spectral efficiency.  This, in turn, has led to an increasingly denser and heterogeneous deployment of wireless infrastructure, where the deployed networks are disparate in various features, e.g., \begin{inparaenum} \item access technology  (RAT), \item coverage area per AP, \item deployed frequency band and bandwidth, and \item backhaul capabilities\end{inparaenum}.  As a result, most of the user equipment (UE) in a dense wireless network would be located in the overlapping coverage areas of multiple APs/RATs. UEs with ability to aggregate traffic from  multiple radio links or RATs (e.g. LTE,  WLAN, $5$G) can, thus, leverage multi-link aggregation to boost their throughput and  quality of service (QoS).

Algorithms and techniques to optimally leverage such multi-link  aggregation would be crucial for boosting both the peak rates as well as the area spectral efficiency  in the next generation ($5$G)  wireless networks \cite{And5G14}.  In fact, such architectures are being  standardized for distributing traffic across cellular (LTE) and wireless LAN (WLAN) HetNets through LTE-WLAN Aggregation (LWA)\cite{3gpp_r2156737,4glwa} in LTE Release 13. Similar framework is also in place for dual-connectivity and traffic aggregation across an anchor, e.g. macrocell,  and booster, e.g. small cell, in LTE  \cite{3gpp_ts36425,JhaIntel14,Zak13}. Note that these architectural frameworks not only allow dynamic traffic aggregation, but also enable seamless offloading across RATs.

Smart UE to AP association strategies in multi-RAT heterogeneous networks (HetNets)  has attracted significant interest from both academia and industry (see e.g. \cite{kimvecyan,madan2010cell,ye2012user,AndLoadCommag13} and references therein). Most of these works, however, do not leverage   UE's  multi-link aggregation capability,  and dynamic distribution of traffic and resource allocation across RATs is not accounted for.  Techniques to realize  capacity gains enabled by such architectures not only need to leverage the added capacity in the network, but do so while ensuring system wide fairness and minimal impact to legacy UEs, i.e. those without multi-link aggregation capability.  These issues have been partially investigated in \cite{Muk14,SinCL15}. The work in \cite{Muk14} analyzed traffic aggregation from a single UE/flow perspective, but system wide fairness was  not  captured. In \cite{SinCL15}, a proportional fair solution for traffic splitting was proposed, where the resource allocation was optimized only at the macrocell or anchor node. This paper, on the other hand, proposes a solution for maximizing a general network utility,  while optimizing the resource allocation across all RATs/APs in the system. The solutions developed in \cite{Sou08,Bethan14}, in the context of multi-band aggregation and joint transmission in massive MIMO networks respectively, are not directly applicable to the HetNet setting  of  this paper.  In particular, the work in \cite{Sou08} investigated proportional fair allocation for carrier aggregation, whereas in this work we consider a more general $\alpha$ fairness (of which proportional fairness is a special case) framework for multi-link aggregation.  The work in \cite{Bethan14} investigated simultaneous connectivity in massive MIMO networks, where the transmission were not orthogonal, whereas in the multi-RAT setting of this work, orthogonal transmissions enable low complexity decentralized solutions converging to optimal solutions.    

In this paper, we propose and demonstrate an algorithm for traffic splitting and aggregation in HetNets, where each UE's traffic is split across multiple RATs/APs. These RATs/APs could constitute macrocells, smallcells, wireless LAN (WLAN) APs, etc. and hence this solution is applicable to LWA and dual connectivity architectures. The proposed algorithm  maximizes a general network wide utility, which incorporates maximizing sum rate, maximizing minimum rate, and  proportional fairness as its special cases. The proposed solution takes into account each UE's spectral efficiency on the available RATs and is amenable to both a centralized and decentralized implementation. The decentralized implementation is realized through a series of message exchanges between the RATs and UEs. The developed framework, thus, generalizes the optimal association solutions for HetNets (like \cite{kimvecyan,ye2012user}).  Furthermore, a closed form solution is developed for the scenarios like LWA, where UE's traffic is split across at most two RATs.   Using comprehensive LTE-WLAN based simulations, the throughput  and capacity gains from the proposed aggregation algorithm are shown to be upto $70\%$ over the baseline single  RAT association algorithms.

\section{System Model and Problem Formulation}\label{sec:sysmodel}
\begin{figure}
  \centering
{\includegraphics[width= \columnwidth]{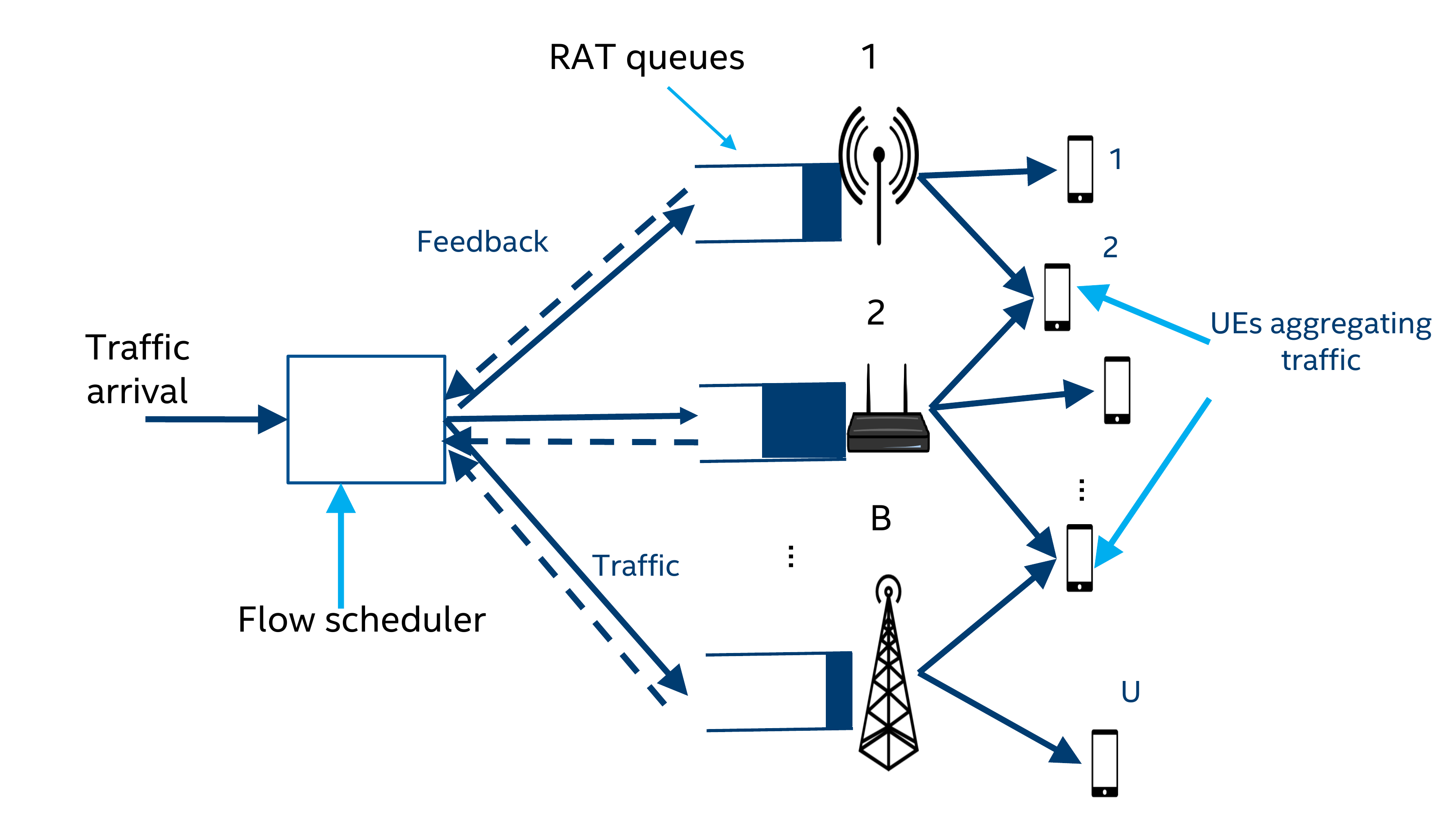}}
\caption{Overview of system architecture where the flow scheduler distributes the flow for UEs across multiple RATs/APs.}
 \label{fig:sysmodel}
\end{figure}

A heterogeneous  multi-RAT network   is considered, where the total number $\NUE$ of active, i.e. with downlink traffic, UEs in the system  can associate with a subset of the total number $\NBS$ of APs operating on non-overlapping frequency bands. These $\NBS$ APs could constitute LTE smallcells, macrocells, WLAN APs, etc. as shown in Fig. \ref{fig:sysmodel}.    All the UEs are assumed to have the capability to aggregate traffic over multiple APs/RATs. The peak capacity for UE $u$ on BS $b$ is denoted by $\pse_{u,b}$  and fraction of  resources allocated to user $u$ on RAT $b$ is denoted by  $\af_{u,b}$, resulting in rate of  $\af_{u,b}\pse_{u,b}$ for UE $u$ on RAT $b$ and cumulative throughput of $\sum_{b}\af_{u,b}\pse_{u,b}$ for UE $u$. Note that $\NUE$ denotes the number of active UEs in a certain part of the network, which dynamically varies over time.

The traffic/flow for all the $\NUE$ UEs enter the network at the flow scheduler (as in Fig. \ref{fig:sysmodel}) and appropriate portion of the traffic for each UE is routed through each RAT/BS\footnote{For example, this flow scheduler is located at Packet Data Convergence Protocol (PDCP) layer in LTE eNodeB for LWA \cite{4glwa}.}. It is assumed that these portions are proportional to the UE's throughput on the corresponding RATs, i.e., the traffic for UE $u$ is split in the ratio $\af_{u,1}\pse_{u,1}$:$\af_{u,2}\pse_{u,2}$:$\ldots$:$\af_{u,\NBS}\pse_{u,\NBS}$. The flow scheduler determines the appropriate resource fractions $\pmb{\af}$ using the feedback of $\pmb{\pse}$ obtained from the respective RATs. Note that the such feedback can be made available through the established mechanisms in aforementioned architectures like  LWA and dual-connectivity \cite{3gpp_r2156967,3gpp_ts36425}. The peak rates $\pmb \pse$ is  assumed to be known at the RATs using channel quality indicator (CQI) or channel state information (CSI) feedback. The notation used in this paper is summarized in Table \ref{tbl:notationtable}.

 \begin{table}
	\centering
\caption{Notation}
	\label{tbl:notationtable}
  \begin{tabulary}{\columnwidth}{ |C | L |}
    \hline
  \textbf{ Notation} & \textbf{Parameter} \\\hline
 $\util_\alpha$, $\alpha$ & Parameterized utility, degree of fairness   \\\hline
 $\res_{u,b} $ & Fraction of resources allocated to UE $u$ on RAT $b$ \\\hline
 $\pse_{u,b}$ & Peak rate of UE $u$ on RAT $b$  \\\hline
 $\rate_u$ & Total throughput of UE $u$ \\\hline
 $\NBS$, $\NUE$ & Number of RATs, number of active UEs respectively  \\ \hline
\end{tabulary}
\end{table}

The traffic distribution and aggregation optimization problem is formulated as a network utility maximization (NUM) as follows
\begin{subequations}\label{eq:opt}
\begin{align}
\text{maximize } & \sum_{u=1}^\NUE \util_{\alpha}(\rate_u) \nonumber \\
\text{subject to } & r_u \leq \sum_{b=1}^\NBS \res_{u,b}\pse_{u,b}\,\, \forall u=1\ldots \NUE  \label{eq:optc1}\\
& \sum_{u=1}^\NUE \af_{u,b} = 1 \,\,\forall b=1\ldots \NBS \label{eq:optc2} \\
& 0\leq\res_{u,b}\leq 1\,\,\forall u=1\ldots \NUE \,\,\forall b=1\ldots \NBS,  \label{eq:optc3} 
\end{align}
\end{subequations}
where $\util_{\alpha}(\rate_u)$ is a utility of UE $u$'s throughput $\rate_u$. The utility is  parameterized by $\alpha$, which specifies the degree of fairness across users. 
The constraint (\ref{eq:optc1})  implies that the user throughput is bounded by the sum of individual rates across all RATs. Constraint (\ref{eq:optc2}) specifies that the resource allocation fraction at each RAT sum to one. The fractional nature of $\res_{u,b}$ is captured in constraint (\ref{eq:optc3}). The $\alpha$-optimal utility function is given by 
$$ \util_{\alpha}(x)  =\begin{cases}
\log(x), \text{ $\alpha$=1}\\
\frac{x^{1-\alpha}}{1-\alpha}, \text{ otherwise.}\\
\end{cases}$$
The above utility function supports a range of objectives like
\begin{itemize}
\item $\alpha=0$: $\max \sum_{u=1}^\NUE \util_{\alpha}(\rate_u) = \max \sum_{u=1}^\NUE \rate_u$, i.e.,  maximize sum rate.
\item  $\alpha=1$: $\max \sum_{u=1}^\NUE \util_{\alpha}(\rate_u) = \max \sum_{u=1}^\NUE \log(\rate_u)$, i.e., maximize  sum log rate or proportional fairness (PF). 
\item  $\alpha\to \infty$: $\max \sum_{u=1}^\NUE \util_{\alpha}(\rate_u) = \max \min_{u=1}^\NUE \rate_u$, i.e., maximize  minimum rate.
\end{itemize}
A similar $\alpha$-optimal formulation was employed in \cite{kimvecyan}, except that the utility was expressed in terms of mean delay for each UE. Note that the solution for (\ref{eq:opt})  leads to a generalization for RAT selection  strategies (e.g. \cite{kimvecyan,ye2012user}), by addressing joint resource allocation and association, as the UE $u$ is assigned only to those RATs, where $\af^*_{u,b}>0$ ($\af^*_{u,b}$ being the optimal solution of (\ref{eq:opt})).

\section{ Aggregation Solution}
The problem posed in (\ref{eq:opt}) is convex and thus guarantees an optimal resource allocation solution, which can be obtained using a convex solver. However, the computational complexity of such an implementation could be considerable given that the state of wireless network is time varying. Therefore, this section analyzes the  Lagrange dual problem of (\ref{eq:opt}), which is similar to \cite{Bethan14}. However, the absence of orthogonality constraints leads to a low complexity solution in this setup.

The Lagrangian function for the problem (\ref{eq:opt}) with the dual variables $\pmb \ratel$ for constraint (\ref{eq:optc1}) and $\pmb \loadl$ for constraint (\ref{eq:optc2})  is  
\small
\begin{align}
&L( \pmb{\res,\rate,\ratel,\loadl})\nonumber \\&=  \sum_{u=1}^\NUE\util_{\alpha}(\rate_u) - \sum_{u=1}^\NUE \ratel_u(\rate_u - \sum_{b=1}^\NBS \res_{u,b}\pse_{u,b}) - \sum_{b=1}^\NBS \loadl_b \left(\sum_{u=1}^\NUE \res_{u,b}-1\right)\nonumber\\
&= \sum_{u=1}^\NUE\util_{\alpha}(\rate_u) -\ratel_u \rate_u + \sum_{b=1}^\NBS \loadl_b +\sum_{u=1}^\NUE\sum_{b=1}^\NBS \res_{u,b}\left(\ratel_u\pse_{u,b}-\loadl_b\right).\label{eq:lag}
\end{align}
\normalsize
The dual function is given by maximizing the Lagragian function over the primal variables or 
\begin{equation}\label{eq:dfn}
D(\pmb{\loadl,\ratel}) \triangleq \max_{\pmb{\res, \rate} \geq 0} L(\pmb{\res,\rate,\ratel,\loadl})
\end{equation}
and the dual program is given by 
\begin{align*}
 \text{minimize }  & D(\pmb{\loadl,\ratel})\\
\text{subject to }  &  \pmb{ \loadl,\ratel} \geq 0.
\end{align*}
Using (\ref{eq:lag}) and (\ref{eq:dfn}), it is noted that $D(\pmb{\loadl,\ratel}) = \infty$ if $\ratel_u\pse_{u,b}-\loadl_b>0$ for some $(u,b)$.
When $\ratel_u\pse_{u,b}-\loadl_b\leq 0\,\, \forall (u,b)$;  $L(\pmb{\res,\rate,\ratel,\loadl})$ is maximized with  $\sum_{u,b} \res_{u,b} (\ratel_u\pse_{u,b}-\loadl_b) \to 0$. Thus, the dual program is simplified as 
\begin{subequations}
\begin{align*}
 \min_{\pmb {\loadl,\ratel}}  \max_{\pmb r} &\sum_{u=1}^\NUE \util_{\alpha}(\rate_u)- \ratel_u \rate_u  + \sum_{b=1}^\NBS \loadl_b   \\
\text{subject to } & \ratel_u\pse_{u,b}\leq \loadl_b \,\,\forall (u,b)   \\
&  \pmb{ \ratel, \loadl, r} \geq 0. 
\end{align*}
\end{subequations}
 Maximizing over $\bf r$ gives 
\begin{equation*}
 \max_{\pmb r} \sum_u \util_\alpha(\rate_u) - \ratel_u \rate_u  =
 \begin{cases}
 \sum_u\frac{1}{\rho-1}\ratel_u ^{1-\rho} \text { for } \rho\neq 1 \\
 \sum_u -1-\log(\ratel_u)  \text { for } \rho =  1  
\end{cases}
\end{equation*}
 where $\rho = 1/\alpha$.  
 
 For $\alpha=1$ (PF), the dual program is 
\begin{subequations} 
\begin{align*}
 \min_{\pmb{\loadl,\ratel}} & \sum_{b=1}^\NBS \loadl_b - \sum_{u=1}^\NUE \log(\ratel_u)  \\
\text{subject to } & \ratel_u\leq \frac{\loadl_b}{\pse_{u,b}} \,\,\forall (u,b)   \\
& \pmb{\ratel, \loadl} \geq 0,
\end{align*}
\end{subequations}
 
or \begin{subequations} 
\begin{align*}
 \min_{\pmb \loadl} & \sum_{b=1}^\NBS \loadl_b - \sum_{u=1}^\NUE \log\left(\min_{b \in \{1 \ldots \NBS \}} \frac{\loadl_b}{\pse_{u,b}}\right)    \\
\text{subject to } &  {\pmb \loadl} \geq 0.  \\
\end{align*}
\end{subequations}
\setcounter{equation}{3}
If $B_u \triangleq \argmax_{b \in \{1 \ldots \NBS \}} \frac{\pse_{u,b}}{\loadl_b}$, where ties are broken arbitrarily, and $ U_b \triangleq \left\{u \in \{1 \ldots \NUE \}\text{ s.t. } B_u = b\right\}$, the dual program is 
\begin{subequations} \label{eq:dualpgm2}
\begin{align}
 \min_{\pmb \loadl}F \triangleq \min_{\pmb \loadl} & \sum_{b=1}^\NBS \loadl_b - \sum_{b=1}^\NBS \sum_{u\in U_b}\log\frac{\loadl_b}{\pse_{u,b}}   \\
\text{subject to } & \pmb{ \loadl} \geq 0.  
\end{align}
\end{subequations}

For $\alpha\neq 1$, the dual program is 
\begin{subequations} 
\begin{align*}
 \min_{\pmb {\loadl,\ratel}} & \sum_{b=1}^\NBS \loadl_b + \sum_{u=1}^\NUE \frac{1}{\rho-1} \ratel_u^{1-\rho}  \\
\text{subject to } & \ratel_u\leq \frac{\loadl_b}{\pse_{u,b}} \,\,\forall (u,b)   \\
& \pmb {\ratel, \loadl} \geq 0, 
\end{align*}
\end{subequations}
 \setcounter{equation}{4}
 or \begin{subequations} \label{eq:dualpgm}
\begin{align}
 \min_{\pmb \loadl}F \triangleq \min_{\pmb \loadl}  & \sum_{b=1}^\NBS \loadl_b + \frac{1}{\rho-1} \sum_{b=1}^\NBS \sum_{u \in U_b} \left(\frac{\loadl_b}{\pse_{u,b}}\right)^{1-\rho}   \\
\text{subject to } &  \pmb \loadl \geq 0. 
\end{align}
\end{subequations}
\begin{rem}\label{rem:KKT} The KKT conditions for the dual program are given by 
\begin{align}\label{eq:KKT1}
\sum_{b\in B^*_u}\res^*_{u,b} \pse_{u,b} & = \max_{b \in \{1\ldots \NBS\}} \left(\frac{\pse_{u,b}}{\loadl_b^*}\right)^{1/\alpha} \forall u,
\end{align} 
where  \begin{align}\label{eq:KKT2}
 B^*_u = \left\{ j:  \frac{\pse_{u,j}}{\loadl_j^*} = \argmax_{b \in \{1\ldots \NBS\}} \frac{\pse_{u,b}}{\loadl_b^*}\right\}
 \end{align}  and  $\pmb \loadl^*$ is the optimal dual variables for (\ref{eq:dualpgm2}) and (\ref{eq:dualpgm}).
\end{rem}
There are two key observations from the above remark: \begin{inparaenum} \item UEs associate  with only those RATs, which have the largest ratio of peak rate to optimal dual variable (as seen from (\ref{eq:KKT2})); and \item thus, the dual variable $\lambda_b$ can be interpreted as the ``load indicator" on RAT $b$ and hence the ratio $\pse_{u,b}/\lambda_b$ as the ``rate indicator" for UE $u$ on RAT $b$. 
\end{inparaenum}

The  dual program in (\ref{eq:dualpgm2}) and (\ref{eq:dualpgm}) can be solved by a subgradient descent approach outlined in Algorithm \ref{optalgo}. 
\begin{algorithm}
\caption{Optimal load indicators}
\label{optalgo}
\begin{algorithmic}[1]
\Procedure{OPT-LOAD}{}
\State Each BS is assigned with initial load indicators $\loadl_{b,1}$. 
\State  $i=1$
\While {$i \leq {N}$}
\State Evaluate $U_{b,i}$ based on the current load indicators.
\State Compute the subgradient for each load indicator $$\nabla F(\loadl_b) = 1-\sum_{U_{b,i}}\frac{\pse_{u,b}^{\rho-1}}{\loadl_{b,i}^\rho}\,\, \forall b.$$
\State Update the load indicator for each BS as 
$$\loadl_{b,i+1} = \loadl_{b,i} + \epsilon_i \left( \sum_{U_{b,i}}\frac{\pse_{u,b}^{\rho-1}}{\loadl_{b,i}^\rho}-1\right) \,\, \forall b. $$
\State $i=i+1$
\EndWhile
\EndProcedure
\end{algorithmic}
\end{algorithm}
\begin{prop} The divergence of the best solution found up till $N$ iterations of  Algorithm \ref{optalgo} from the optimal dual program objective  $F^*$ is  
\begin{equation}
\|F_{\mathrm{best}}^{N} - F ^*\| \leq \frac{(\pmb\loadl_{.,1}- \pmb \loadl^*)^2 + G^2\sum_{i=1}^N \epsilon_i^2}{2\sum_{i=1}^N\epsilon_i},
\end{equation}
where $G$ upper bounds the norm of subgradients, i.e., $\|\nabla F(\pmb\loadl)\|_2\leq G$, and $\pmb \loadl^*$ are the optimal load indicators.
\end{prop}
\begin{IEEEproof}
See \cite{boyd2006subgradient}.
\end{IEEEproof}

\begin{figure}
  \centering
{\includegraphics[width= \columnwidth]{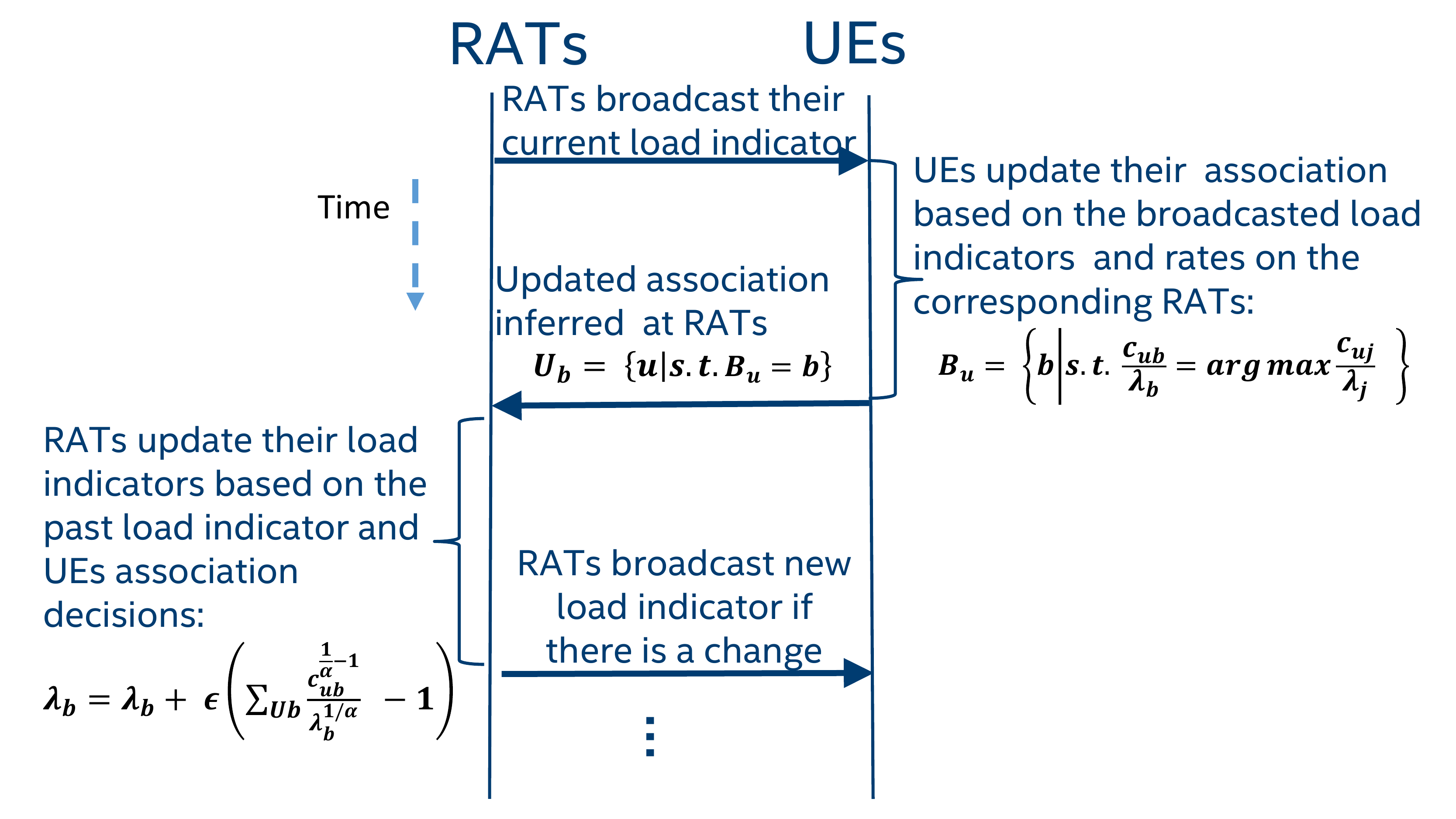}}
\caption{Decentralized implementation of Algorithm \ref{optalgo}.}
 \label{fig:deflow}
\end{figure}

The algorithm for estimating the optimal load indicators is  amenable to a decentralized implementation, where the optimal solution  is derived through an iterative  sequence of message exchanges  between the network  and the UEs. The steps $5$-$7$ of Algorithm \ref{optalgo} are illustrated in Fig. \ref{fig:deflow}.

Using Remark \ref{rem:KKT}, the optimal resource fractions for UEs associating with a single RAT are
\begin{equation}\label{eq:res1}
\res_{u,B^*_u}^*=\frac{\pse_{u,B^*_u}^{\rho-1}}{\loadl_{B^*_u}^{*\rho}}\,\, \forall u \,\,\mathrm{s.t.}\,\, |B^*_u|=1.
\end{equation}
For the rest of UEs, the optimal resource fractions can be found by solving the following set of equations:
\begin{subequations}\label{eq:KKTsplit}
\begin{align}
\sum_{b\in B^*_u}\res_{u,b}^* \pse_{u,b} & = \max_b \left(\frac{\pse_{u,b}}{\loadl_b^*}\right)^{1/\alpha} \forall u \in \{1 \ldots \NUE\}\\
\sum_{u} \res_{u,b}^* &=1 \,\, \forall b \in \{1 \ldots \NBS\} .
\end{align} 
\end{subequations}
The solution for  (\ref{eq:KKTsplit}) is available in closed form for certain plausible scenarios enabled by the following theorem. 

\begin{thm}\label{thm:nuesplit}
Under $\alpha$-optimal NUM, among the UEs having connectivity  to $M$ RATs no more than $M-1$ UEs  should split traffic across those $M$ BSs/RATs at any instant.
\end{thm}
\begin{IEEEproof}
The loop-removal procedure proof in \cite{Sou08} for $\alpha=1$ can be extended to a general $\alpha$-optimal utility by replacing the KKT conditions as 
\begin{align*}
T_u \triangleq \sum_{b}\res_{u,b} \pse_{u,b} & = \max_b \left(\frac{\pse_{u,b}}{\loadl_b}\right)^{1/\alpha} \forall u.
\end{align*} 
\end{IEEEproof}

As per the above theorem, in the scenario where a set of users aggregate among only two RATs simultaneously like LTE and WLAN, no more than one UE's traffic in that set is simultaneously transmitted from both the RATs. As a result, the corresponding resource fractions for the UE $u$ that aggregates is obtained by simplifying (\ref{eq:KKTsplit}) as
\begin{equation}\label{eq:res2}
\res_{u,b}^* =1 -\sum_{u\in U_b} \res_{u,b}^*  \,\, \forall b \in B^*_u ,\,\, \forall u \,\,\mathrm{s.t.}\,\, |B^*_u|=2.
\end{equation}
Therefore,  for the case of two RAT aggregation, like LWA or dual-connectivity,  the optimal resource allocation fractions are developed in closed form in (\ref{eq:res1}) and (\ref{eq:res2}).

\section{Performance Evaluation}
\begin{figure}
  \centering
{\includegraphics[width= \columnwidth]{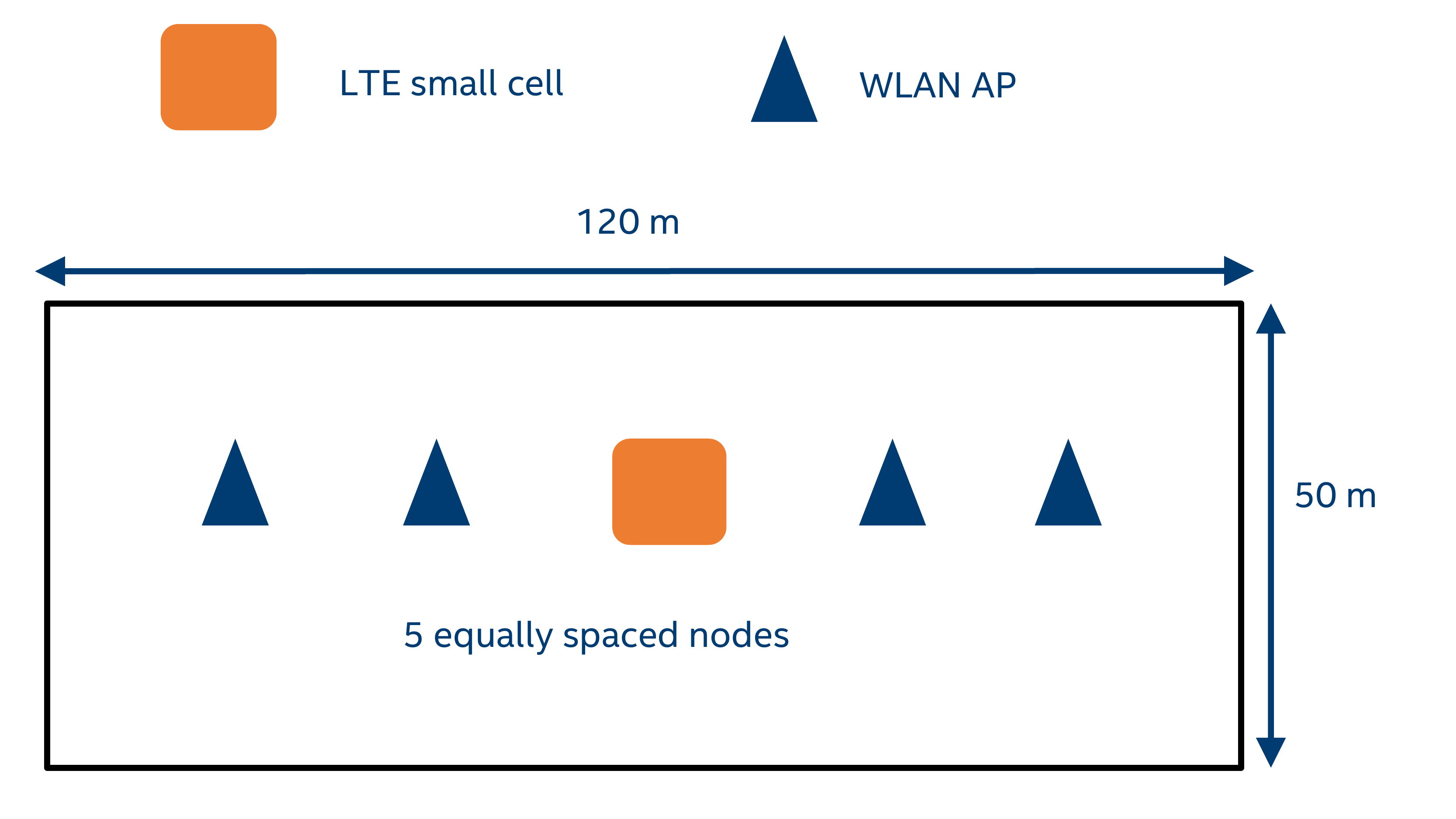}}
\caption{Deployment of LTE small cell and WLAN APs in the building. }
 \label{fig:deployment}
\end{figure}

The developed algorithm is evaluated using a comprehensive system level LTE-WLAN HetNet simulator. This investigation is also aimed at assessing  the possible performance gains from LWA framework  over existing multi-RAT inter-working solutions. The simulation parameters and  assumptions are listed in Table \ref{tbl:param}. An enterprise like deployment scenario is considered in this setup, where each building contains one LTE small cell and four non-collocated  WLAN APs (see Fig. \ref{fig:deployment}). The UEs in the building aggregate traffic over their WLAN AP and the LTE small cell of the building.  This scenario is particularly attractive for LWA deployment. The proposed NUM based algorithm (termed ``LWA-NUM" henceforth) is compared with two RAT association algorithms: \begin{inparaenum} \item LTE Release 12 (Rel12) and \item Release 13 based radio inter-working solutions (Rel13) \cite{4glwa}\end{inparaenum}.  In Rel12, a UE associates with the  WLAN AP only when the $\SINR$ from LTE macrocell is below a certain threshold and $\SNR$ from WLAN AP is above a certain threshold.  The optimal value for these thresholds is empirically found for the presented results.  In Rel13 based solution, upon arrival of a new file/flow for a UE, it is routed  through the  RAT (LTE or WLAN) that provides higher throughput to that UE. Since multiple files are downloaded during the simulations, each UE may thus associate with different RATs for different file download during each simulation trial. However, no dynamic traffic distribution across RATs is allowed within a file download.  Therefore, on the one hand, the Rel12 based optimized association is static, but benefits from optimal thresholds found through exhaustive simulations, whereas on the other hand, the Rel13 solution is dynamic and reactive to current network conditions, while following a ``greedy" per-user strategy. Note that these RAT selection algorithms also use the ``local" resource allocation algorithms at each AP based on the corresponding schedulers (as per Table \ref{tbl:param}).

 Note that both the Rel12 and Rel13 algorithms lead to UEs associating with either LTE or WLAN,  and no traffic splitting/aggregation is done, whereas with LWA-NUM the traffic is distributed dynamically across the two RATs based on the current network conditions, while incorporating for  multi-user fairness.  For the presented results, $\alpha=1$ is used leading to  proportional fairness based solution. The load in the network is varied by varying the mean inter-arrival time for files, where the size of file is fixed as per Table \ref{tbl:param}.  The  inter-arrival rate is varied as  $0.3$s, $0.4$s, and $0.6$s, which lead to approximately $20\%$, $40\%$, and $60\%$ network utilization  respectively in our setup.

\begin{figure}
  \centering
{\includegraphics[width= \columnwidth]{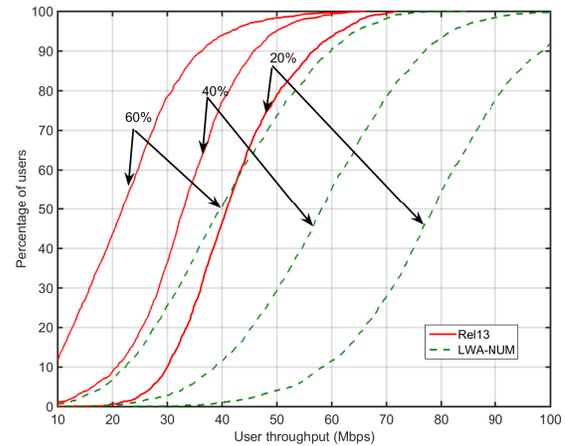}}
\caption{Comparison of user throughput CDF  obtained from  Rel13 and the proposed algorithm  for  different  network utilization levels. }
 \label{fig:tputcomp}
\end{figure}

The distribution of per UE throughput in the network obtained from the  simulations  is shown in Fig.~\ref{fig:tputcomp} for  LWA-NUM and Rel13 association algorithm with LTE-WLAN backhaul latency of $5$ ms. As can been seen, for all the range of network utilization levels, the proposed algorithm outperforms and stochastically dominates the Rel13 based solution for all UEs. In particular,  the proposed algorithm provides the same median rate ($50$ percentile rate) with $60\%$ network utilization as that provided by Rel13 with $20\%$ utilization, which in turn  implies a $3$x gain from the perspective of load supported while delivering the same QoS.

\begin{figure}
  \centering
{\includegraphics[width= \columnwidth]{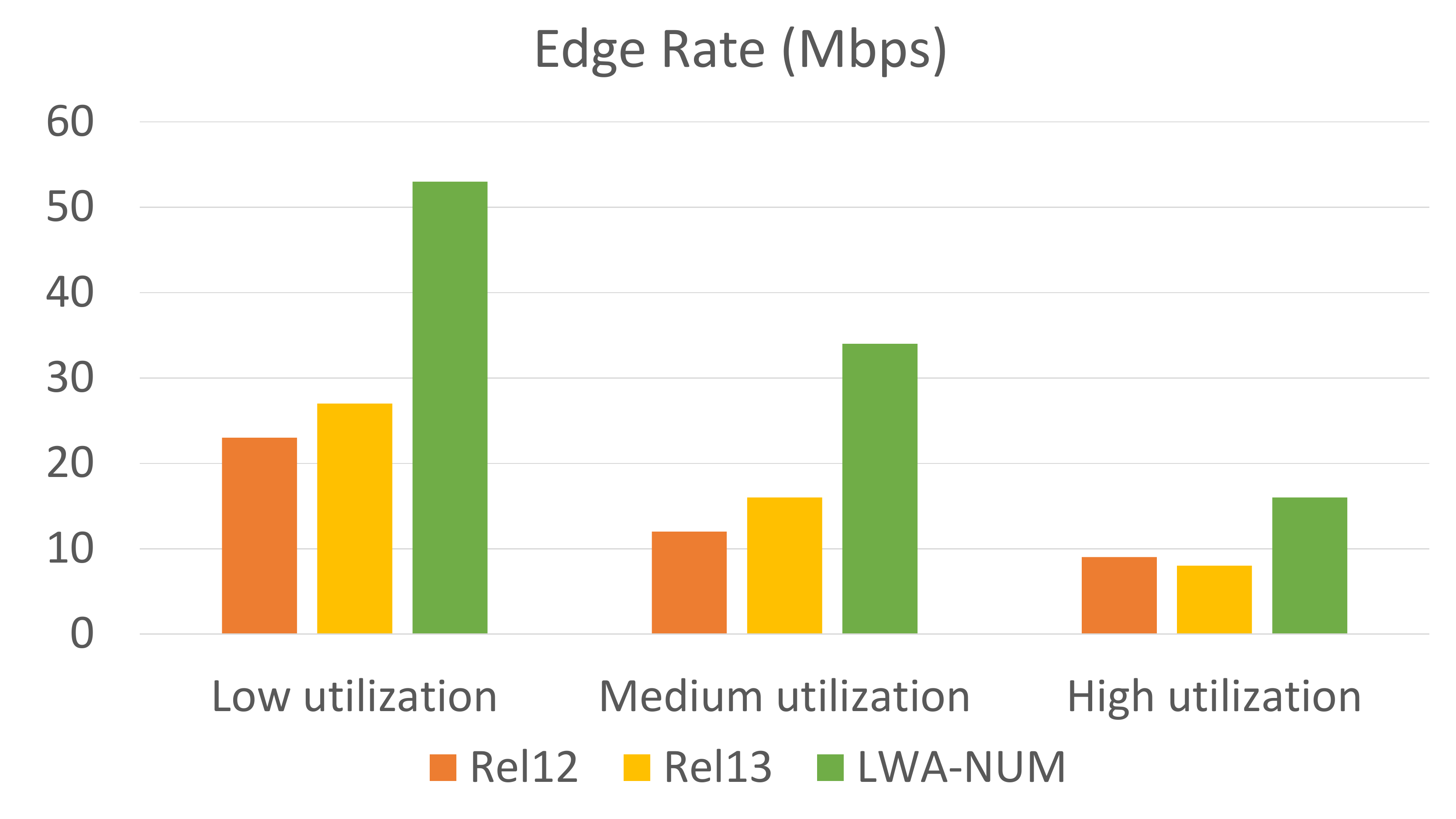}}
\caption{Fifth percentile throughput from  different algorithms for various network load.}
 \label{fig:edgerate}
\end{figure}

\begin{figure}
  \centering
{\includegraphics[width= \columnwidth]{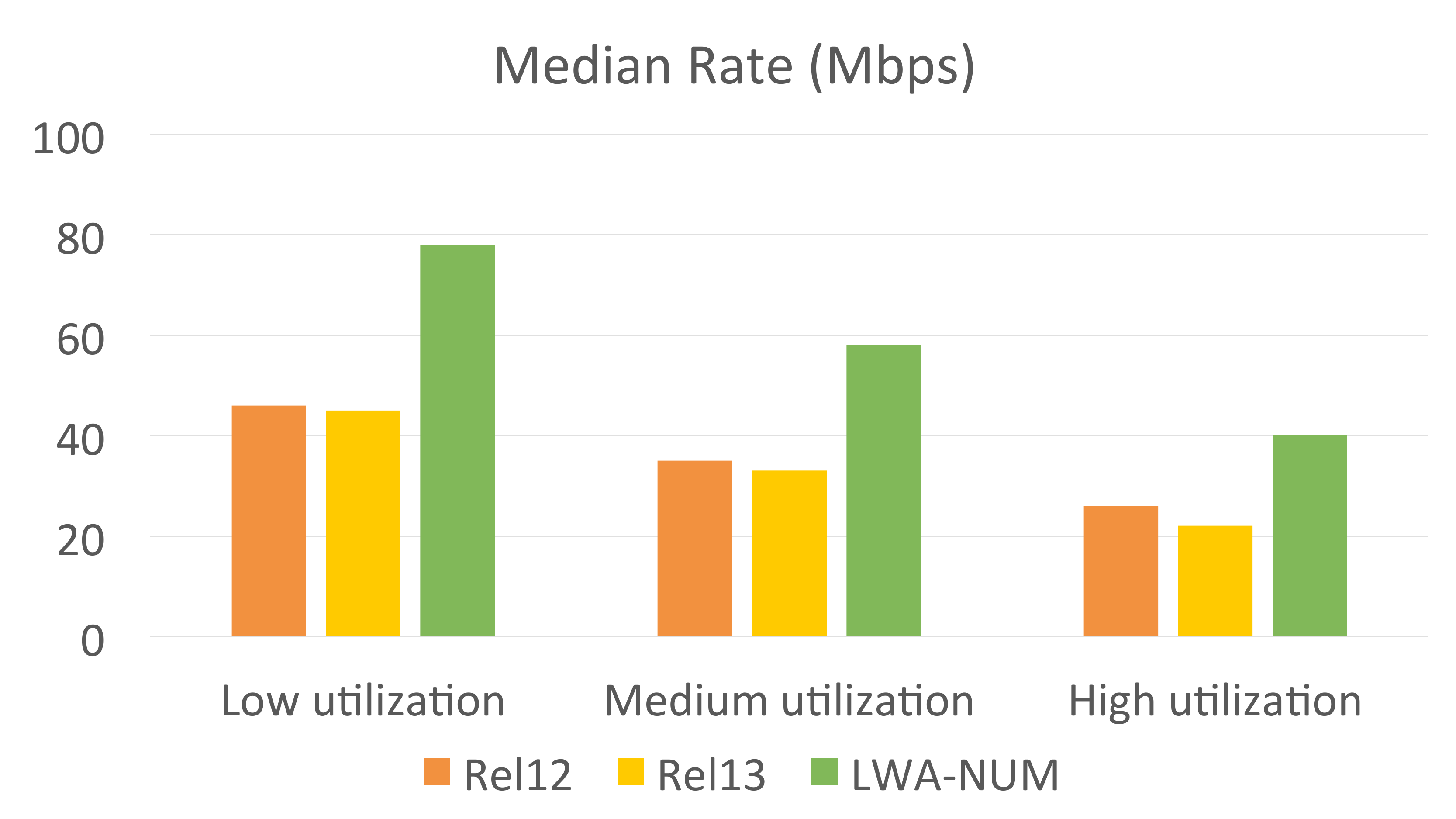}}
\caption{Median throughput from  different algorithms for various network load.}
 \label{fig:medianrate}
\end{figure}

 The fifth percentile (or edge)  rate   and median rates obtained from the three algorithms are shown in Fig. \ref{fig:edgerate} and \ref{fig:medianrate} respectively for different network utilization levels. There are two key  observations to be derived from these two figures: \begin{inparaenum} \item Rel13 based RAT association performs similar to Rel12 in median rates and provides marginal gain in edge rates; and \item  the proposed algorithm provides about $180\%$ gain in edge  rates and $70\%$ gain in median rates over the association algorithms. \end{inparaenum} The justification for the first observation is the fact that the thresholds for Rel12 are empirically optimized, however in a realistic setting Rel13 based association is more attractive. The second observation stems from the fact that LWA-NUM  allows for dynamic steering of traffic from one RAT to another as network conditions evolve (even within file downloads) as enabled by LWA framework \cite{3gpp_r2156737},  while incorporating for multi-user fairness. 
 

\begin{table}
	\centering
\caption{Simulation Assumptions}
	\label{tbl:param}
  \begin{tabulary}{\linewidth}{|C|C| }
    \hline
    \textbf{Parameter} & \textbf{Description} \\\hline
    Topology &	3 sectors per macrocell,  7 cell wrap-around,  1 building per sector with 1 LTE small cell, 4 WLAN APs and 10 UEs uniformly distributed inside a building
  \\\hline
Channel model & Indoor model in TR 36.889 Annex 1  (ITU InH channel model) \\\hline
Traffic & Downlink non full buffer with exponentially distributed variable average inter-arrival  time, and fixed file size of $0.5$ MB/file across all UEs (3GPP FTP traffic model 3) \\\hline
Carrier frequency  & $2$ GHz (LTE), $2.4$ GHz (WLAN 802.11n)  \\\hline
Bandwidth & $10$ MHz (LTE), $20$ MHz (WLAN) \\\hline
Transmit power & $24$ dBm (LTE smallcell), $18$ dBm (WLAN AP)\\\hline
No. antennas (macro, pico, WLAN, UE)	 & ($2$, $2$, $2$, $2$) \\\hline
Antenna configuration	& macrocell, smallcell: co-polarized, UE: co-polarized    \\\hline
Max rank per UE	& $2$ (SU-MIMO)  \\\hline
UE channel estimation	& Ideal  \\\hline
Feedback/control channel errors &	No Error  \\\hline
PHY abstraction &	Mutual information (LTE),  RBIR (WLAN) \\\hline
LTE scheduler, scheduling granularity	& Proportional fair,  $5$ PRBs \\\hline
Receiver type	& Interference unaware MMSE  \\\hline
Feedback periodicity	& $10$ms  \\\hline
CQI and PMI feedback granularity  in frequency &	5 PRBs  \\\hline
PMI feedback	& 3GPP Rel-10 LTE codebook (per sub-band)  \\\hline
Outer loop for target FER control &	$10\%$ PER for 1st transmission  \\\hline
Link adaptation	& MCSs based on LTE transport format  \\\hline
HARQ scheme &	CC  \\\hline
    WLAN TXOP &  $1$ ms \\\hline
    WLAN MIMO mode & Closed loop (w/o feedback overhead) \\\hline
WLAN MPDU Size&	$1500$ Bytes  \\\hline
LTE-WLAN backhaul latency & $5$ms \\\hline
\end{tabulary}
\end{table}

\section{Conclusion}
This paper presents a framework for optimal  traffic aggregation solution in wireless multi-RAT HetNets. The presented work generalizes the load balancing and user assignment solutions studied in past.  To the authors' best knowledge, this is the first work to propose and demonstrate an $\alpha$-optimal algorithm for aggregating flows with LWA and demonstrate gains over RAT selection techniques.  The developed framework is also applicable to  RATs employing  millimeter-wave based frequencies. Future work could investigate (perhaps extending  stochastic geometry based analysis e.g. \cite{SinDhiAnd13}) the operating regimes, where traffic aggregation is expected to yield maximum gains over just association/selection schemes.

\input{SplitDraftCR.bbl}
\end{document}

%% file: notation.tex
\DeclareMathOperator*{\argmax}{arg\,max}

\newcommand{\rate}{r}

\newcommand{\pse}{c}

\newcommand{\res}{\eta}
\newcommand{\loadl}{\lambda}
\newcommand{\ratel}{\nu}
\newcommand{\util}{f}
\newcommand{\NUE}{\mathrm{U}}
\newcommand{\NBS}{\mathrm{B}}

\newcommand{\SINR}{\mathtt{SINR}}

\newcommand{\SNR}{\mathtt{SNR}}

\newcommand{\af}{\eta} 
 
\newtheorem{thm}{{\bf Theorem}}

\newtheorem{rem}{Remark}

\newtheorem{prop}{Proposition}

\theoremstyle{definition}

\theoremstyle{thm}